\documentclass{iaus}
\usepackage{graphicx}

\title[Worldwide R\&D of VO] 
{Worldwide R\&D of Virtual Observatory}

\author[Cui \& Zhao]   
{Chenzhou Cui$^1$,Yongheng Zhao$^1$}

\affiliation{$^1$National Astronomical Observatories, Chinese
Academy of Sciences, Beijing 100012, China \break email:
ccz@bao.ac.cn}

\pubyear{2008}
\volume{xxx}  
\pagerange{119--126}
\date{?? and in revised form ??}
\jname{Proceedings Title IAU Symposium}
\editors{A.C. Editor, B.D. Editor \& C.E. Editor, eds.}
\begin{document}

\maketitle

\begin{abstract}
Virtual Observatory (VO) is a data intensive online astronomical
research and education environment, taking advantages of advanced
information technologies to achieve seamless and uniform access to
astronomical information. The concept of VO was introduced in late
of 1990s to meet challenges brought up with data avalanche in
astronomy. This paper reviews current status of International
Virtual Observatory Alliance, technical highlights from world wide
VO projects, and a brief introduction of Chinese Virtual
Observatory.

\end{abstract}

\firstsection 
\section{Introduction}

During the last decade, advances in technologies have been changing
the abilities and ambitions of astronomers. New technologies on
telescope design and fabrication bring more powerful telescopes then
ever to astronomers. Large scale digital sky surveys are prospered
with the appearance of CCD mosaic camera. The scale of numerical
simulation is also increasing rapidly with the development of
hardware and software. As a result of these advances, data avalanche
is occurring in Astronomy. Furthermore, driven by multi-waveband sky
surveys and observations, new astronomical fields appeared and are
becoming more and more popular, for example, multi-waveband
research, multi-archive data mining, time domain analysis, precise
cosmology, etc (\cite{Lawrence2006}). To meet challenges brought up
by the above changes, Virtual Observatory (VO) concept was initiated
(\cite{Szalay2001}). Virtual Observatory is a data intensive online
astronomical research and education environment, taking advantages
of advanced information technologies to achieve seamless and uniform
access to astronomical information. The power of the World Wide Web
is its transparency. It is as if all the documents in the world are
inside your PC. The idea of the VO is to achieve the same
transparency for astronomical data and information
(\cite{Quinn2004}). VO is a science driven but technic enabled
resolution, a cyber-infrastructure for the 21st century astronomy.

\section{International VO activities}

After National Virtual Observatory (US-VO), the first funded VO
project, VO projects were initiated in different countries and
regions. International Virtual Observatory Alliance (IVOA) was
formed in June 2002 with a mission to "facilitate the international
coordination and collaboration necessary for the development and
deployment of the tools, systems and organizational structures
necessary to enable the international utilization of astronomical
archives as an integrated and interoperating virtual observatory."
At present, the IVOA is formed by 16 projects. VO framework includes
agreed standards, interoperable data collections, interoperable
services and applications. IVOA focuses its work on development of
standards. Now, more then 20 specifications have been published by
IVOA working groups.

Following the concept of VO and IVOA specifications, a new
prosperous era is coming for astronomical softwares and services.
New applications, VO-enhanced legacy services are supporting
astronomer's research more and more strongly. Core services from
US-VO provide functions including data discover, service registry,
catalog coverage, object cross match, source extraction and
identification. UK VO project, AstroGrid deployed the world¡¯s only
unified VO operational service "AstroGrid", which gives astronomers
a "one-stop" access to all the worlds¡¯ astronomy data from its
desktop. Many other new applications are release by world-wide VO
projects and contributors. For example, VOPlot, VOSpec, VisIVO,
SAADA, etc. At the same time, existing applications and systems are
upgraded and enhanced to support VO including CDS services (SIMBAD,
VizieR, Aladin), NED, SciSoft, IRAF, TOPCAT, Montage, SExtractor,
and so on.

\section{VO in China}
Chinese Virtual Observatory (China-VO) is the national VO project in
China initiated in 2002 (\cite{Cui2004}). The China-VO aims to
provide VO environments for Chinese astronomers. It focuses its
research and development on applications and VO science in the
following five fields: 1.China-VO Platform, providing VO
environments for Chinese astronomical community; 2.Uniform Access to
Global Astronomical Resources and Services, importing international
resources to Chinese astronomers and sharing Chinese resources to
international colleagues; 3.VO-ready Projects and Facilities,
collaborating with astronomical projects to prompt they are
VO-compliant; 4.VO-based Astronomical Research Activities, guiding
and training astronomers to use VO applications and services;
5.VO-based Public Education, developing non-professional services
for the public.

During the last several years, several VO applications and services
have been initiated and developed by the China-VO. For example,
VOFilter, an XML filter for OpenOffice.org Calc to load VOTable
files; SkyMouse, a smart interface for astronomical on-line
resources and services; FitHAS (FITS Header Archiving System), a
toolkit for FITS file providers and data centers; VO-DAS, an
OGSA-DAI based service system to provide unified access to astronomy
data. Furthermore, the China-VO is tightly collaborating with
LAMOST, a Chinese ambitious spectrum sky survey project, to make it
is VO-enabled, sharing its archives and software to the VO.

Internet is inter-connected. VO is trying to link on-line resources
and services together in a higher level. More and more VO supported
resources and services are available. How to do better science using
these tools? VO is not a simple thing, but a way of life.
Astronomers need learn how to survive in the VO era.

\begin{acknowledgments}
The China-VO is supported by NSFC under contract No. 60603057,
90412016 and 10778623.
\end{acknowledgments}


\begin{thebibliography}{}

\bibitem[Cui (2004)]{Cui2004}
     {Cui, C. \& Zhao, Y.} 2004,
     \textit{PNAOC} 1(3), 203

\bibitem[Lawrence (2006)]{Lawrence2006}
     {Lawrence, A.} 2008,
     in: K.A. van der Hucht (eds.),
     \textit{Highlights of Astronomy},
     The Virtual Observatory in action : new science, new technology, and next generation facilities. vol.\ 14, in print

\bibitem[Quinn (2004)]{Quinn2004}
     {Quinn, P. J., Barnes, D. G., Csabai, I., et al.} 2004,
     in: P. J. Quinn \& A. Bridger (eds.),
     \textit{Proceedings of the SPIE},
     Optimizing Scientific Return for Astronomy through Information Technologies, vol.\ 5493, p.\ 137

\bibitem[Szalay (2001)]{Szalay2001}
     {Szalay, A. \& Gray, J.} 2000,
     \textit{Science} 293, 2037


\end{thebibliography}
\end{document}